# The impact of NFT profile pictures within social network communities

SIMONE CASALE-BRUNET, École Polytechnique Fédérale de Lausanne, Switzerland
MIRKO ZICHICHI, Universidad Politécnica de Madrid, Spain
LEE HUTCHINSON, WhaleAnalytica.com, Switzerland
MARCO MATTAVELLI, École Polytechnique Fédérale de Lausanne, Switzerland
STEFANO FERRETTI, University of Urbino "Carlo Bo", Italy

This paper presents an analysis of the role of social media, specifically Twitter, in the context of non-fungible tokens, better known as NFTs. Such emerging technology framing the creation and exchange of digital object, started years ago with early projects such as "CryptoPunks" and since early 2021, has received an increasing interest by a community of people creating, buying, selling NFTs and by the media reporting to the general public. In this work it is shown how the landscape of one class of projects, specifically those used as social media profile pictures, has become mainstream with leading projects such as "Bored Ape Yacht Club", "Cool Cats" and "Doodles". This work illustrates how heterogeneous data was collected from the Ethereum blockchain and Twitter and then analysed using algorithms and state-of-art metrics related to graphs. The initial results show that from a social network perspective, the collections of most popular NFTs can be considered as a single community around NFTs. Thus, while each project has its own value and volume of exchange, on a social level all of them are primarily influenced by the evolution of values and trades of "Bored Ape Yacht Club" collection.

CCS Concepts: • **Applied computing → Sociology**.

Additional Key Words and Phrases: NFT, PFP, Twitter, blockchain, Ethereum, profile picture, CryptoPunks, Bored Ape Yacht Club

## 1 INTRODUCTION

Non Fungible Tokens (NFTs) are a booming technology, with a trading volume that - in just one year - has increased from $150 million to over $5 billion (recorded in January 2022 by considering only the transactions that took place on the Ethereum blockchain [2] through the main exchange platform OpenSea) [19]. Public attention towards NFTs has exploded since the beginning of 2021, when the NFT market experienced record sales and exchange volumes, up to the present day where individual pieces for some projects have reached prices of millions of dollars and have been purchased by a number of celebrities. However, little is known about the overall structure and evolution of the NFT ecosystem, or how this phenomenon has also grown thanks to the communities that have formed in social networks, such as Twitter. In a simplistic way, we can define an NFT as a unique digital asset whose information (such as features, traits and ownership) is certified and managed using blockchain technology. In other words, we can think of a NFT project as a set of football player cards. Each card of the set is different from the others because of some specific traits (e.g., a player can be more or less wanted), the circulation of each card is fixed, and the owner of each card is known and traceable. The set of cards is managed on the blockchain by means of a smart contract, while each card represents an indivisible token. More specifically, NFTs are an example of a Distributed Ledger Technology (DLT) application that renders trustworthy intermediaries such as banks and notary firms obsolete [24]. In fact, cryptographic tokens based on DLTs, through guaranteed enforcement of encoded obligations (e.g., through smart contracts), can combine both concepts of (i) access rights to an underlying economic value (property) [5, 13], and (ii) a permission to access someone else's property or services or a collective good. NFTs and their technology are already being used in a number of varied fields. Some examples include: the certification of digital generative artworks, the representation and management of land and objects ownership in the metaverse, and the certification of ownership of a profile image in social networks, such as Twitter. It is this latter use case of profile picture images (PFPs) that has been one of the largest contributors to the popularity of NFT technology during the past year. There have been digital figurine projects whose cards were previously trading for a few hundred dollars in early 2021 and whose average price today is in the hundreds of thousands of dollars. Leaving aside the purely economic aspect, this phenomenon of using NFTs as profile pictures has both social and technological interest that, in our opinion, are worth researching deeper. In fact, the use of profile pictures belonging to a digital collection with a limited circulation (typically 10,000 different figurines) and whose ownership is certified by a distributed ledger, introduces some entirely new and disruptive concepts. First of all, those who use a profile picture, with no apparent relation to their own photo, seek to eliminate stereotypes and preconceptions by hiding themselves in a sort of anonymity typical of the crypto world; moreover, it allows the user the ability to create a digital persona (that may or may not have any relation to their real identity from which their subconscious may wish to escape) [26]. Second, the fact of having a limited number of cards (a limit forced by construction by the smart contract written on blockchain) provides the possession of some of them (such as the CryptoPunks and Bored Ape Yacht Club) with real digital status symbols that allow the owner to highlight the fact that she or he has been a person in the crypto world for a long time or economically successful. Third, the owner of a figurine in a particular project can sometimes enjoy digital or real benefits: such as being able to access restricted discussion rooms in social networks like Discord, or take part real world or metaverse special events and parties. This allows a union of real and digital worlds where the concept of ownership is easily transportable and usable, without requiring going through intermediaries. From the personal experience of the authors of this work, Twitter and Discord represent the two most representative and used social networks for people interested in NFTs of the PFPs type, where real communities have been created. This work, the first of its kind, aims to analyze the activity of the most PFPs in the Twitter ecosystem: which turns out to be both the main communication channel of these communities and the one most easily analyzed thanks to its content query API. In this paper we wanted to apply state of art social network analysis (SNA) techniques to analyze these communities and verify how they interact and drive the NFTs market and how they are interconnected. SNA techniques are in fact widely used in application areas such as community and topic detection [6]. In this paper we took 18 most significant NFT profile pictures projects and collected all their blockhain transactions and all their related tweets starting from the date of creation to April 15th, 2022. Next, we analyzed the data obtained from Twitter by creating two different kind of graphs: one related to tweets with commonalities between each project pair and a second related to tweets having common hashtags. We then analyzed which communities are the most influential in this ecosystem and how they interact with each other. The paper is structured as follows: Section 2 presents an overview of SNA techniques available in literature. Next, Section 3 discusses which projects have been selected for this work, the rationale for choosing them among others and how the data were collected from Twitter and the Ethereum blockchain.

Next, Section 4 presents which SNA analyses have been conducted and the results obtained. Finally, Section Section 5 concludes this paper by providing future research insights.

## 2 STATE OF ART

Over the past decades, social media data have been studied to understand people [1, 3] and exploited to develop strategies for promotion, prediction and engagement in different fields [8, 11, 25]. A growing body of existing research has analysed the dissemination of information online, investigating how buying products or joining communities is affected by the mechanics of social networks [20]. A considerable amount of work in this area has been done on sentiment analysis and opinion mining using Twitter. This is due to the fact that Twitter data is mostly composed of text, i.e. tweets. Twitter is largely popular for its institutional role of showing news and reports, but also renowned as the medium where users freely share their true feelings, write what they are doing and discuss a wide range of topics, which include places, people and products [11]. The purpose of sentiment analysis, thus, is to understand a user's opinions and attitudes about different topics through the texts they have written [15]. Even if generally the understanding of communities in social network is performed through sentiment analysis and text mining [15], in this work we will focus on Social Network Analysis (SNA). In social networks, different mechanisms generate various structures such as friendship networks, mention networks, hashtag networks, etc. Generally, Twitter users use hashtags to add context and metadata to tweets making twitter more expressive [20]. Hashtags, indeed, provide a way to search for any kind of content posted by any kind of user, e.g. personal feelings, public criticism, nonsense messages, important updates [11]. Based on these, it is possible to create hashtag networks in Twitter by converting the co-occurrences of these tags in a single tweet into links. In several cases, these types of networks exhibit the same universal properties as real networks, such as the small-world property, clustering and power-law distributions [11, 21]. The majority of research in this context focuses on the dissemination of information and the analysis of the networks of interactions formed among users around different topics, such as popular television broadcasts [11], health care [14], COVID-19 pandemics [10]. This kind of study relates to the SNA that govern the relationships between NFTs, their creators and their collectors as described in [23] where the auction dynamics of NFTs, links between artists and collectors and co-bidding networks were studied, concluding strong evidence of first-mover advantage. Other studies were conducted using data obtained from DLT or marketplaces. In the context of NFTs transacions, the study of how different wallets interact was performed in [7] In this study, an attempt was made to identify the wallets that most influence trades using graph-based analysis techniques. Preliminary studies between the relationship between the average price of tokens in an NFT collection and twitter trends was analysed in [17] and [18]. Both works applied learning techniques to predict the future price of assets. Although the datasets considered are in both cases limited and restricted to specific NFT collections, it is possible to conclude that social activity may be a very important feature to predict over time the collection price.

## 3 PROJECTS SELECTION AND DATA COLLECTION

For this study, we selected 18 different profile picture NFTs projects that, following our experience in the field, we found to be significant each one for a specific reason. These projects are illustrated in Table 1 where reported for each one are: the publishing date of the smart contract on the blockchain (also called deployment date), the Ethereum address of the smart contract, the official Twitter account, and the list of hashtags that we identified as the ones most used by the respective Twitter community. CryptoPunks (Figure 1a) are commonly regarded as the originators of PFP projects: they were launched in June 2017 and consist of 10,000 punks that are all different and pixelized. The community is predominantly composed of people who have been in the crypto scene for a long time (referred to in the crypto slang as OGs). We included The HashMasks (Figure 1b), while considered more fine-art than PFP, because they are a prototypical collection launched just before the so-called NFT bull run of 2021-2022. This collection introduced the centrality of the concept of traits and rarity in each image. This concept was then the basis for all the projects we analyzed in this paper. We considered projects that became very famous such as the Bored Ape Yacht Club (Figure 1c), further referred as BAYC that are now considered as a status symbol, projects that were born in the same period such as the Crypto Hodlers (Figure 1f) that struggled to expand their social community, projects like the mfers (Figure 1p), created by a long-time crypto investor and Twitter influencer Sartoshi, that has been the forerunner of the usage of the CC0 licence and meme culture. In general, we tried to enclose in this selection all those projects that introduced specific innovations or because their overall volume of tweets and transactions is representative for the study. It is important to point out that this list does not represent any kind of financial advice in any way.

Table 1. Profile picture NFT projects that we analysed: name, smart contract release date, number of issued tokens, Ethereum smart contract address, official Twitter account and the list of most used hashtags by the project community.

| Project | Deployment | Assets | Smart Contract Address | Twitter Account | Hashtags |
|---:|---|---|---|---|---|
| CryptoPunks | 22.06.2017 | 10,000 | 0xb47e3cd837ddf8e4c57f05d70ab865de6e193bbb | @cryptopunksnfts | cryptopunk(s) |
| The HashMasks | 28.01.2021 | 16,384 | 0xc2c747e0f7004f9e8817db2ca4997657a7746928 | @TheHashmasks | HashMask(s), TheHashMask(s), Hashies |
| Bored Ape Yacht Club | 22.04.2021 | 10,000 | 0xbc4ca0eda7647a8ab7c2061c2e118a18a936f13d | @BoredApeYC | BoredApeYachtClub, BoredApe(s), BAYC, ape(s)followape(s), BoredApeYC |
| Meebits | 03.05.2021 | 20,000 | 0x7bd29408f11d2bfc23c34f18275bbf23bb716bc7 | @cryptopunksnfts | meebit(s) |
| Cool Cats | 07.07.2021 | 3,138 | 0x1a92f7381b9f03921564a437210bb9396471050c | @coolcatsnft | coolcatsnft, Cat(s)FollowCat(s), coolcat(s) |
| Crypto Hodlers | 07.07.2021 | 10,000 | 0xe12a2a0fb3fb5089a498386a734df7060c1693b8 | @Hodlers_NFT | Hodlers_NFT, HodlersNFT |
| Gutter Cat | 07.07.2021 | 3,138 | 0xedb61f74b0d09b2558f1eeb79b247c1f363ae452 | @GutterCatGang | GutterCatGang |
| The Alien Boy | 12.07.2021 | 10,000 | 0x4581649af66bccaee81eebae3ddc0511fe4c5312 | @TheAlienBoyNFT | TheAlienBoyNFT |
| World of Women | 27.07.2021 | 10,000 | 0xe785e82358879f061bc3dcac6f0444462d4b5330 | @worldofwomennft | worldofwomennft |
| DeadFellaz | 13.08.2021 | 10,000 | 0x2acab3dea77832c09420663b0e1cb386031ba17b | @Deadfellaz | Deadfellaz, Deadfella(s) |
| 0N1 Force | 15.08.2021 | 7,777 | 0x3bf2922f4520a8ba0c2efc3d2a1539678dad5e9d | @0n1Force | 0n1Force |
| Creature World | 28.08.2021 | 10,000 | 0xc92ceddfb8dd984a89fb494c376f9a48b999aafc | @creatureNFT | creatureNFT |
| CryptoMories | 07.10.2021 | 10,000 | 0x1a2f71468f656e97c2f86541e57189f59951efe7 | @CryptoMories | CryptoMorie(s), FaMorie, Tsunamorie, TsunamorieGames |
| MekeVerse | 07.10.2021 | 8,8888 | 0x9a534628b4062e123ce7ee2222ec20b86e16ca8f | @MekaVerse | MekaVerse |
| doodles | 16.10.2021 | 10,000 | 0x8a90cab2b38dba80c64b7734e58ee1db38b8992e | @doodles | DoodlesNFT, @doodles |
| mfers | 29.11.2021 | 10,021 | 0x79fcdef22feed20eddacbb2587640e45491b757f | @sartoshi_nft | Mfer, MfersUnited, sartoshi_nft |
| Alien Frens | 16.12.2021 | 10,000 | 0x123b30e25973fecd8354dd5f41cc45a3065ef88c | @Alienfrens | alienfren(s)nft, alienfren(s) |
| Azuki | 10.01.2022 | 10,000 | 0xed5af388653567af2f388e6224dc7c4b3241c544 | @AzukiOfficial | AzukiOfficial, azuki |

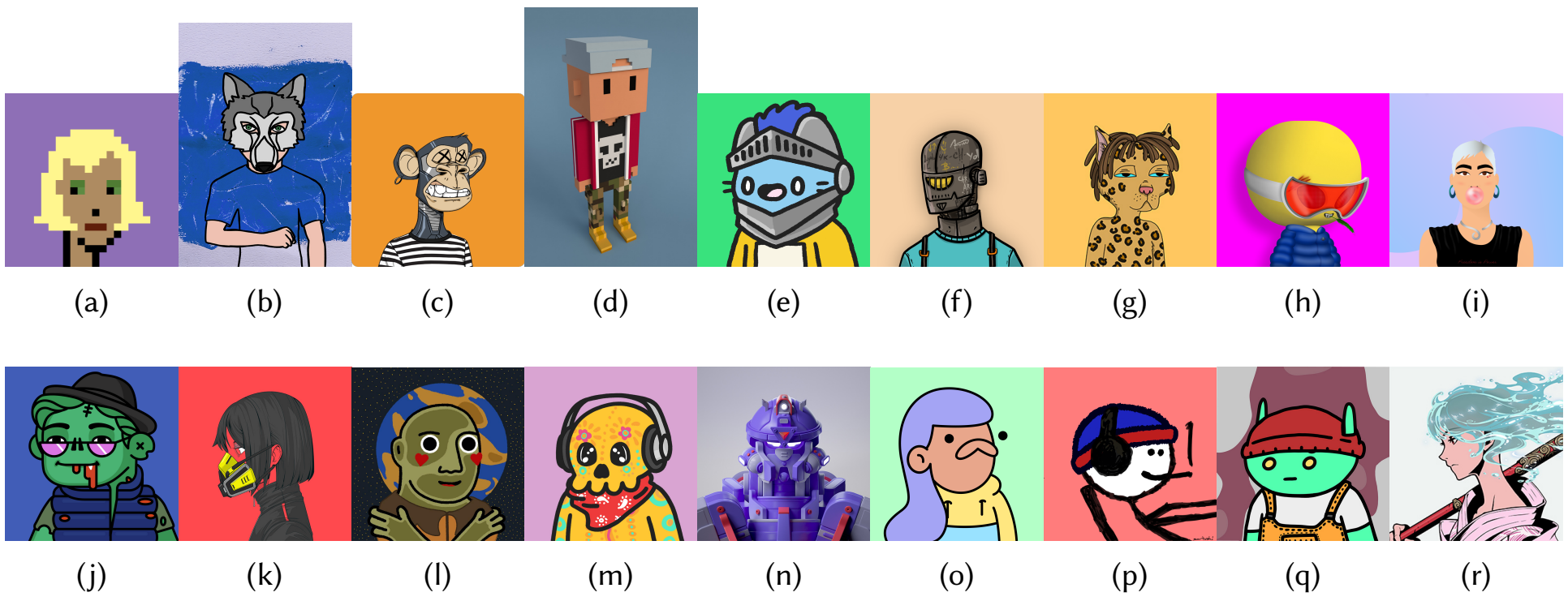

Fig. 1. Profile picture NFT projects that we analysed: (a) CryptoPunks, (b) The HashMasks, (c) Bored Ape Yacht Club, (d) Meebits, (e) Cool Cats, (f) Crypto Hodlers, (g), Gutter Cat Gang, (h) The Alien Boy, (i) World of Women, (j) DeadFellaz, (k) 0N1 Force, (l) Creature, (m) CryptoMories, (n) MekaVerse, (o) doodles, (p) mfer, (q) Alien Frens, (r) Azuki.

Table 2. Projects dataset summary: data has been harvested from the date of their creation to April 15th, 2022.

| | Volume | | | Transactions | | Tweets | |
|---|---|---|---|---|---|---|---|
| Project | ETH | USD | % | Num. | % | Num. | % |
| CryptoPunks | 620,031.08 | 1,752,749,583.99 | 38.48 | 18,588 | 4.70 | 664,366 | 8.58 |
| The HashMasks | 40,930.74 | 83,239,880.08 | 2.54 | 35,323 | 8.93 | 69,326 | 0.89 |
| Bored Ape Yacht Club | 277,928.22 | 941,285,168.64 | 17.25 | 33,455 | 8.46 | 3,061,178 | 39.51 |
| Meebits | 84,964.82 | 268,888,857.28 | 5.27 | 23,625 | 5.97 | 87,785 | 1.13 |
| CoolCats | 92,283.32 | 298,404,993.60 | 5.73 | 26,851 | 6.79 | 449,704 | 5.80 |
| Crypto Hodlers | 809.96 | 1,887,870.11 | 0.05 | 7,698 | 1.95 | 14,043 | 0.18 |
| Gutter Cat Gang | 17,381.66 | 53,049,411.44 | 1.08 | 9,384 | 2.37 | 423,137 | 5.46 |
| The Alien Boy | 2,097.94 | 5,685,271.79 | 0.13 | 15,826 | 4.00 | 115,017 | 1.48 |
| World of Women | 51,441.25 | 164,689,141.78 | 3.19 | 22,512 | 5.69 | 431,586 | 5.57 |
| DeadFellaz | 19,008.37 | 64,879,647.97 | 1.18 | 31,410 | 7.94 | 129,089 | 1.67 |
| 0N1 Force | 38,309.43 | 126,980,644.57 | 2.38 | 19,340 | 4.89 | 180,337 | 2.33 |
| Creature World | 31,513.54 | 107,940,615.40 | 1.96 | 24,612 | 6.22 | 115,467 | 1.49 |
| CryptoMories | 11,745.86 | 41,532,353.10 | 0.73 | 25,152 | 6.36 | 388,013 | 5.01 |
| MekaVerse | 44,552.30 | 162,397,286.41 | 2.76 | 13,327 | 3.37 | 219,593 | 2.83 |
| Doodles | 89,405.03 | 308,007,309.25 | 5.55 | 20,824 | 5.27 | 515,229 | 6.65 |
| mfers | 26,077.74 | 77,737,051.08 | 1.62 | 20,827 | 5.27 | 702,103 | 9.06 |
| Alien Frens | 21,380.50 | 69,619,738.90 | 1.33 | 26,489 | 6.70 | 341,026 | 4.40 |
| Azuki | 141,562.67 | 407,102,715.75 | 8.78 | 20,188 | 5.11 | 269,892 | 3.48 |
| Total | 1,611,424.45 | 4,936,077,541.14 | | 395,431 | | 7,747,078 | |

### 3.1 Blockchain data

The data saved on blockchain, which are public and accessible to all by construction of the Ethereum blockchain, are efficiently accessed when structured in formats such as SQL [4]. For this reason, for the analysis presented in this paper we made use of Google BigQuery which structures this data in an SQL database and provides a very efficient access API [16]. This method of analysis has already been presented in [7] and will be referred to for more technical details about the data extraction techniques used. Briefly, for each transaction we stored the hash of the transaction, the address of the NFT smart contract, the addresses of the wallet(s) or smart contract(s) that sold and purchased an NFT token, the Ethereum block time as milliseconds from the UNIX epoch, the Ethereum block number, the value of ETH and/or WETH (Wrapped ETH) transferred and the number of tokens exchanged during the transaction (can be more than one). Gas fees (i.e., transaction costs) were excluded from the analysis; if the exchange currency is neither ETH nor WETH the transaction was seen as a simple transfer (seen that the vast majority of the transactions happen in ETH or WETH, we concluded that this assumption still produces reasonable estimates of the exchange volumes while significantly reducing the complexity of the queries used). Whenever more than one token was exchanged in the same transaction, the individual price of each token was considered the same, and computed by splitting the total amount of the transaction into equal parts. For each NFT collection, data was extracted from the date of their creation to April 15th, 2022. It must be noted that unlike what was done in [7] where only transactions involving Externally-owed-Accounts (EoAs) were considered, for this work also transactions involving Contract Accounts (CAs) have been considered. In order to eliminate suspicious transactions (e.g., attributable to money laundering activities), for each project we eliminated transactions above the 95th percentile and those where the buyer and seller address are exactly the same. Moreover, only for the Meebits collection we decided to neglect all the transactions made on the LooksRare marketplace because we judged the majority of transactions made within this platform as suspicious: as an example, in mid-January 2021 there were sales of some tokens for millions of dollars each, when the average value traded in the days before was in the tens of thousands of dollars [12]. In Table 2 one can find for each project the number of both Ethereum transactions analysed from each project smart contract deployment date, till April 15th, 2022. The value in USD was calculated using the close spot price value relative to the day of the transaction, and the price provided by [9] has been used. The total number of transactions acquired and analysed in this work is 393,210.

### 3.2 Twitter data

The method that we used to collect tweets from Twitter has been going through the official Search Tweets API [22]. For research purposes, this API allows to programmatically access public tweets from the complete archive dating back to the first Tweet in March 2006, using specific search queries. For each tweet we stored the tweet time as milliseconds from the UNIX epoch, the unique tweet identifier generated by Twitter, the tweet text, the number of replies, likes and retweet count at the time of the query, the list of hashtags used on the tweet, and the unique Twitter username of the author of the tweet. For each project, we queried the API by providing as search points the project official twitter account and specific hashtags used by the project social community (see Table 1). For the scope of this work, only English-language tweets have been collected. In Table 2 one can find for each project the number of tweets from the first tweet found by the query till April 15th, 2022. The total number of tweets acquired and analysed in this work is 7,747,078.

## 4 DATA ANALYSIS AND RESULTS

### 4.1 Ethereum wallets and Twitter users

The NFT market has seen exponential growth from January 2021 to the present both in terms of volumes traded and users who have purchased at least one NFT [2]. As the first analysis, we analysed the growth that occurred on this ecosystem in terms of number of transactions and tweets, and in terms of number of unique Twitter users and Ethereum wallets who participated in the various project communities. The data reported in Table 2, and summarized in Figure 2, shows how in one year the PFP interest on Twitter went from 50 daily tweets in the beginning of January 2021 to more than 50,000 daily tweets in mid April 2022 by considering only the 18 selected projects. In the same period the number of daily unique Twitter users passed from 50 to more than 20,000 with a peak of 60,000, onboarding during the whole period more than 1 million unique active users. Interestingly, from Figure 2 (c) and (d) both the new and cumulative Twitter users and Ethereum wallets follow the same trend: this means that owning one of these NFTs implies therefore also participating in social activity on Twitter.

### 4.2 The communities

In order to identify the interactions and connections that were formed across the communities of each project, we performed as a first analysis one that aimed to identify common Twitter users and Etehereum wallets. For each collection we defined: (i) the list of Twitter users who have posted at least one tweet attributable to a collection (i.e., mentioning the official account, or using one of the hashtags defined in Table 1), (ii) the list of Ethereum wallets that have purchased at least one of the tokens in the collection. Figure 3a depicts a matrix showing for each cell $(i, j)$ the number of common Twitter users between the project shown in the i-row and the one shown in the j-column. The matrix can be read row by row: for each row i, the total number of users is given by the cell on the diagonal (i,i), the cells on the i-th row are then normalized to this. Similarly, it can be read column by column. The same considerations can be made when considering the Ethereum wallets that are illustrated in Figure 3b. From both of these figures we can see that the projects that dominate in terms of Twitter users and Ethereum wallets are the CryptoPunks, BAYC, Cool Cats, Gutter Cat, World of Women, and Doodles. A second type of analysis that we performed was to analyze the tweets in common between pairs of projects and represent this information as an undirected weighted graph. The graph was constructed by defining a node for each project and defining a weighted arc between projects where the weight. As the weight of each node was assigned the number of tweets and as the weight of each arc the number of common tweets

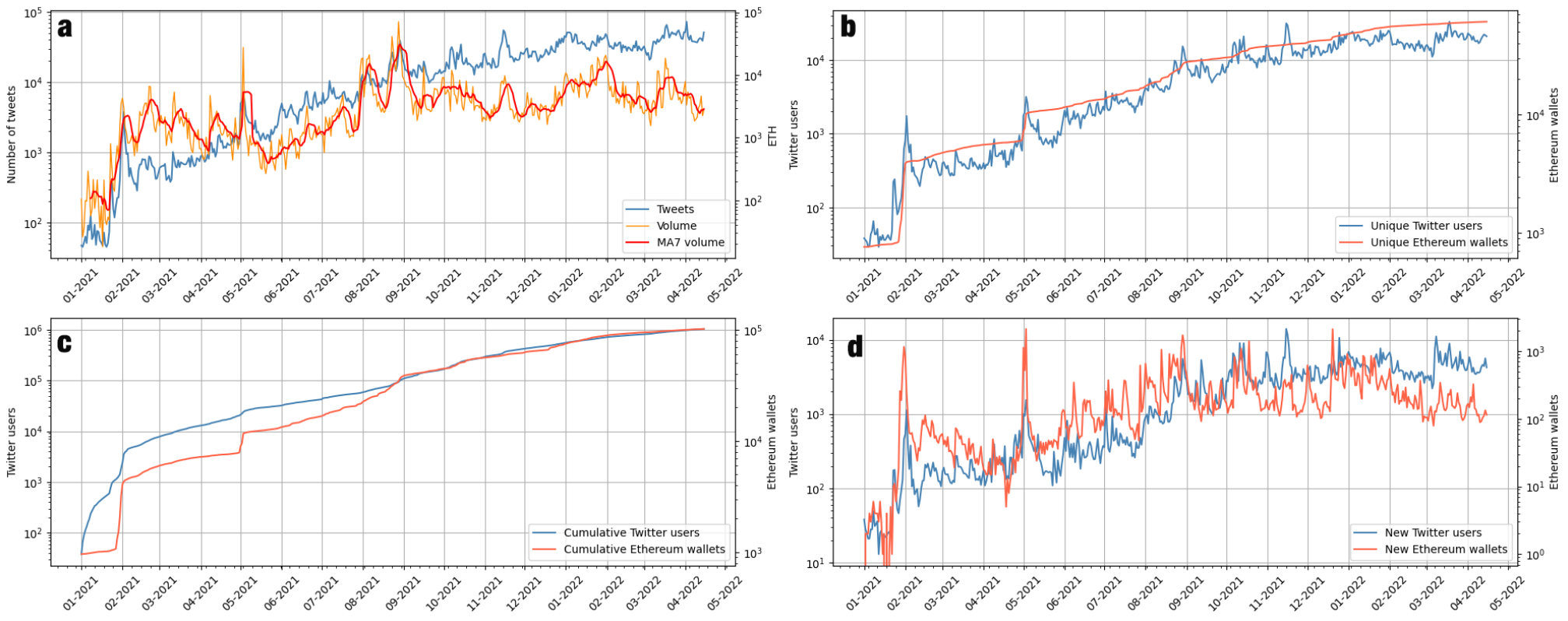

Fig. 2. Twitter and Ethereum volumes and users from January 2021 to April 2022: (a) Tweets and trading volume; (b) Twitter users mentioning the projects and Ethereum wallets owning at least one NFT token from the 18 collections considered; (c) cumulative Twitter users and Ethereum wallets; (d) new Twitter users and new Ethereum wallets owning at least one NFT token from the 18 collections considered.

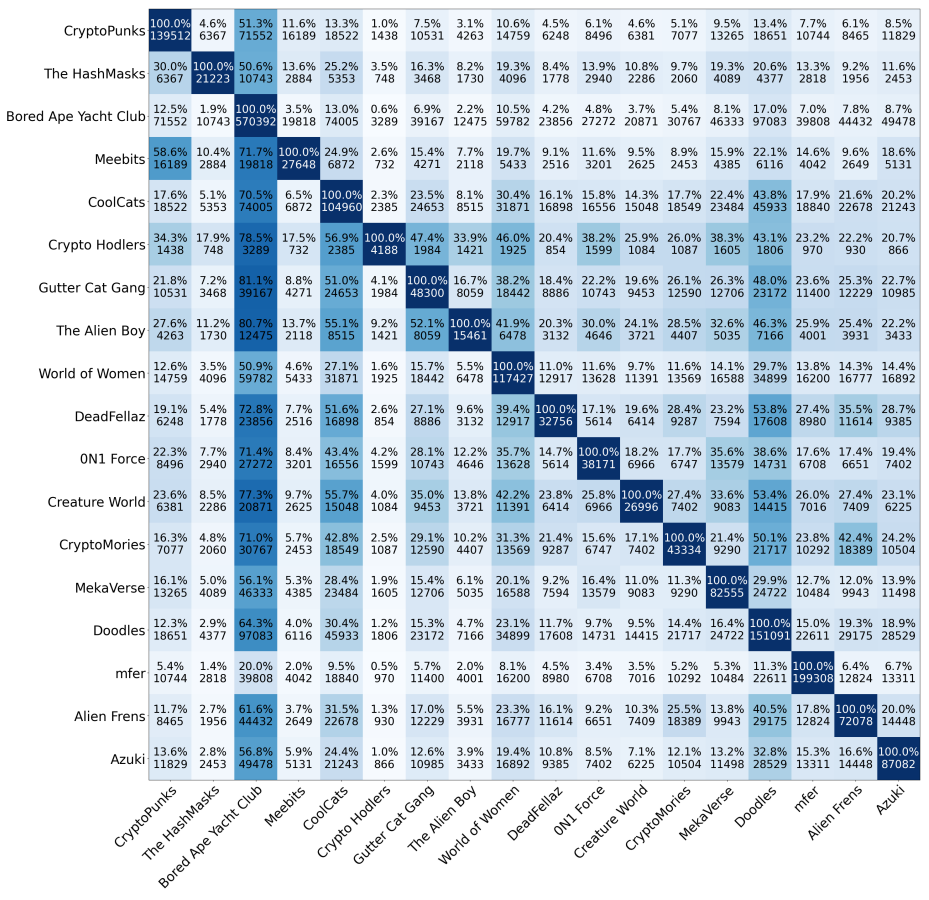

(a) Cumulative Twitter users

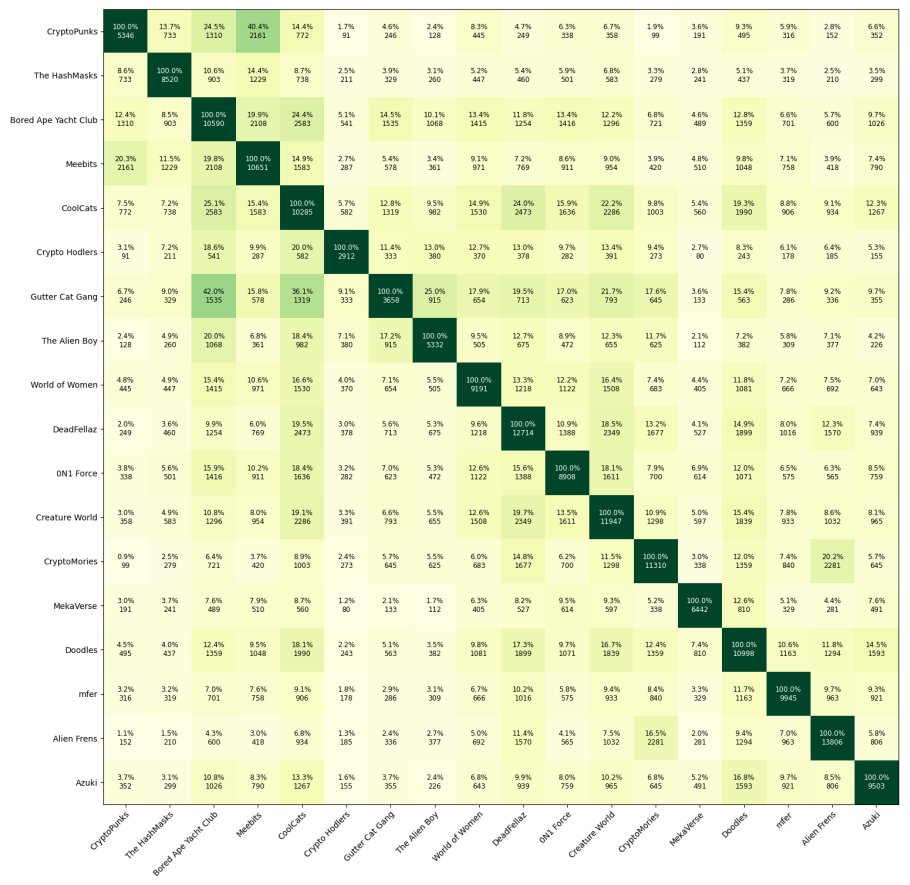

(b) Cumulative Ethereum wallets

Fig. 3. Common (a) Twitter users and (b) Ethereum wallets distribution among the different projects.

(i.e., counting the number of tweets that concern the two projects at the same time). The resulting graph is shown in Figure 4a, while the structural statistics of the graph are summarized in Table 3a. The graph consists of 18 nodes (the number of projects), and 153 arcs in a single connected component. All the nodes are connected with the remaining 17 (see average degree, transitivity and edge density): this indicates how the social community, although polarized toward a few projects, is highly connected (i.e. average clustering equal to 1). For this reason, modularity is equal to 0, i.e. having a single cluster there is no difference among edges within groups and edges expected on the basis of chance [11]. We successively performed a pagerank analysis to identify which nodes were the most influential: results are reported in Figure 5b and we can see that the most influential ones are BAYC, CryptoPunks, Cool Cats, Gutter Cat, World of Women, and Doodles. These are the same ones that were identified in the previous analysis and, as can also be seen from Figure 6a, also the ones with the highest number of associated tweets. For each project we then tried to identify commonality with the others by going to identify the heaviest graph arcs. In other words, for each project we identified the three most related by going for the ones with the most tweets in common. Table 3b summarises these results: you can see that the BAYC are most often present and most connected with the other projects.

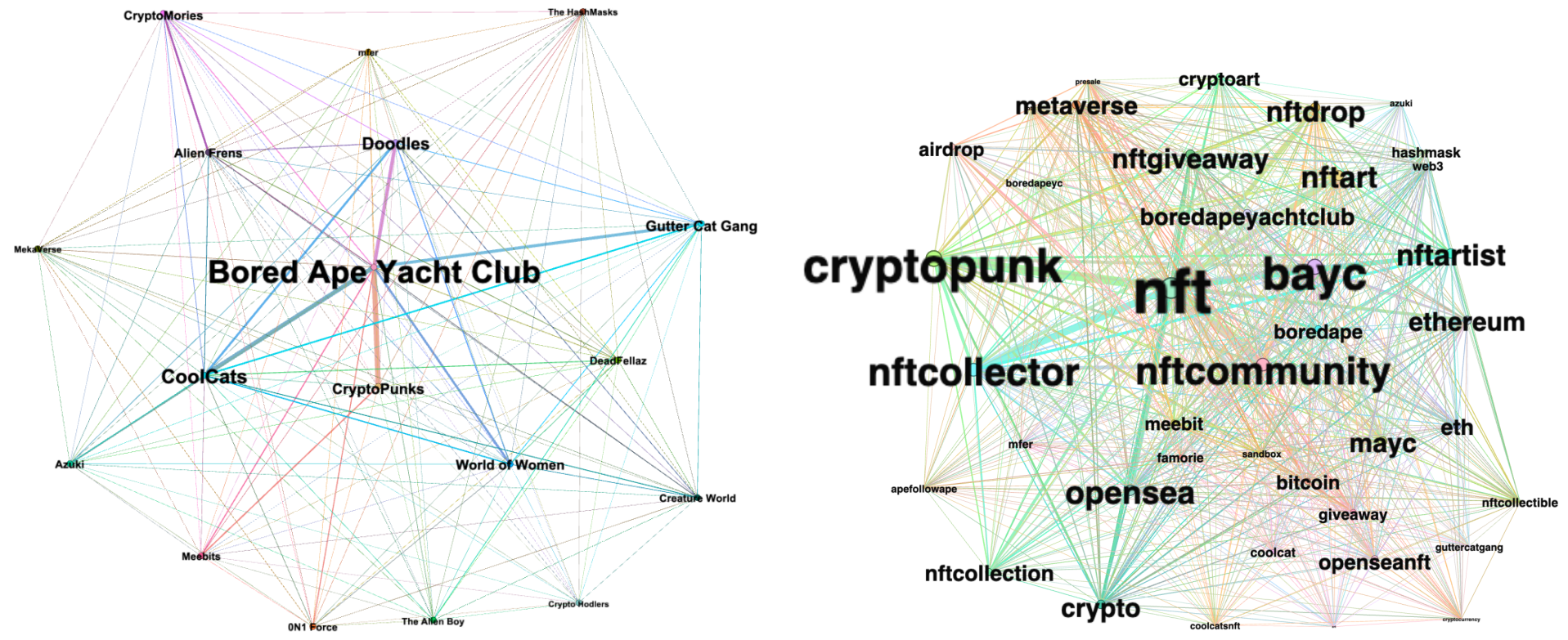

(a) Projects graph: labels proportional to node pagerank. (b) Top 20 hashtags graph: labels proportional to node size.

Fig. 4. Projects and Hashtags graphs where connections represent common tweets.

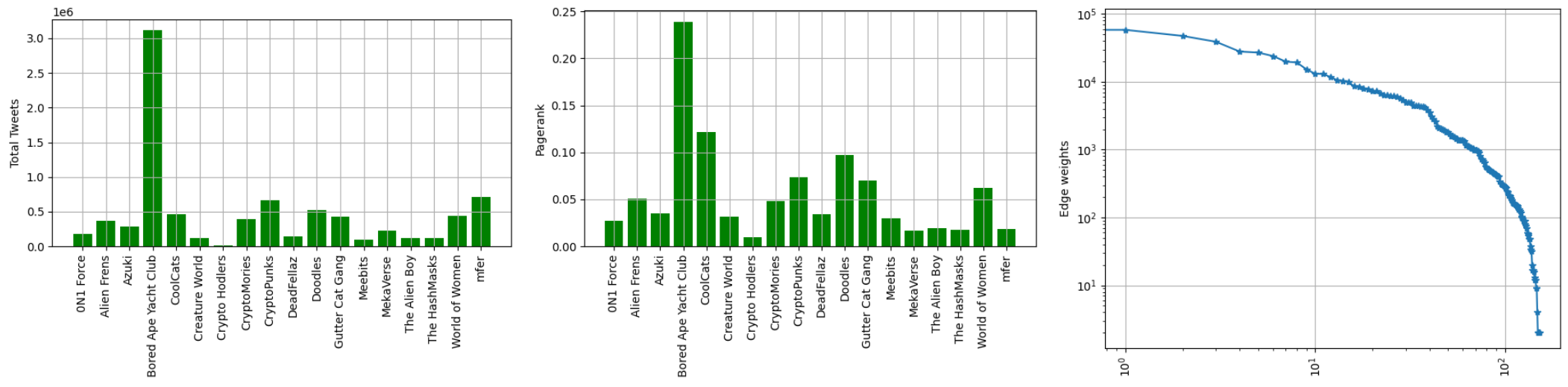

(a) Size (b) Pagerank (c) Nodes weights distribution

Fig. 5. Size, pagerank and node weights distributions of the projects graph illustrated in Table 3.

### 4.3 Hashtags

As a further point of analysis, we focused on hashtags and how they are used in tweets. The hashtags - that are words preceded by the # symbol such as #NFT - are in fact the most common method used by Twitter users to create

Table 3. Projects and Hashtags graph properties and commonality. For this latter, only the top-20 per size are reported.

(a) Graphs main properties

| Property | Projects Graph | Hashtags Graph |
|---|---|---|
| Nodes | 18 | 60060 |
| Edges | 153 | 762950 |
| Transitivity | 1.00 | 0.02 |
| Modularity | 0.00 | 0.05 |
| Average Clustering | 1.00 | 0.80 |
| Edge Density | 1.00 | 0.00 |
| Average Degree | 17.00 | 25.41 |
| Total Triangles | 2448 | 42366009 |
| Average Edge Weights | 4132.72 | 29634.29 |
| Max Edge Weights | 70328 | 627786 |
| Min Edge Weights | 2 | 3 |
| Connected Component | 1 | 439 |

(b) Projects commonality

| Project | Commonality |
|---|---|
| CryptoPunks | Bored Ape Yacht Club, Meebits, CoolCats |
| The HashMasks | Bored Ape Yacht Club, CryptoPunks, Meebits |
| Bored Ape Yacht Club | CryptoPunks, CoolCats, Doodles |
| Meebits | CryptoPunks, Bored Ape Yacht Club, The HashMasks |
| CoolCats | Bored Ape Yacht Club, Doodles, Gutter Cat Gang |
| Crypto Hodlers | Bored Ape Yacht Club, The Alien Boy, World of Women |
| Gutter Cat Gang | Bored Ape Yacht Club, CoolCats, Doodles |
| The Alien Boy | Bored Ape Yacht Club, Gutter Cat Gang, CoolCats |
| World of Women | Bored Ape Yacht Club, CoolCats, Doodles |
| DeadFellaz | CoolCats, Bored Ape Yacht Club, Doodles |
| 0N1 Force | Bored Ape Yacht Club, CoolCats, World of Women |
| Creature World | CoolCats, Bored Ape Yacht Club, Doodles |
| CryptoMories | Alien Frens, Bored Ape Yacht Club, CoolCats |
| MekaVerse | Bored Ape Yacht Club, CoolCats, 0N1 Force |
| Doodles | Bored Ape Yacht Club, CoolCats, World of Women |
| mfer | Bored Ape Yacht Club, Doodles, Azuki |
| Alien Frens | CryptoMories, Bored Ape Yacht Club, Doodles |
| Azuki | Bored Ape Yacht Club, Doodles, CoolCats |

(c) Hashtags commonality

| Hashtag | Commonality |
|---|---|
| nft | cryptopunk, nftcommunity, bayc |
| cryptopunk | nft, bayc, nftcollector |
| bayc | nft, cryptopunk, mayc |
| nftcollector | nft, nftcommunity, cryptopunk |
| nftcommunity | nft, nftcollector, cryptopunk |
| opensea | nft, cryptopunk, bayc |
| nftart | nft, nftcollector, nftcommunity |
| nftartist | nft, nftcollector, nftcommunity |
| nftdrop | nft, nftcollector, nftcommunity |
| nftgiveaway | nft, nftcommunity, nftcollector |
| mayc | bayc, nft, cryptopunk |
| crypto | nft, cryptopunk, bayc |
| ethereum | nft, cryptopunk, bayc |
| metaverse | nft, cryptopunk, bayc |
| boredapeyachtclub | nft, cryptopunk, nftcommunity |
| eth | nft, cryptopunk, bayc |
| nftcollection | nft, nftcollector, nftcommunity |
| openseanft | nft, nftcollector, nftcommunity |
| boredape | nft, nftcommunity, cryptopunk |
| bitcoin | nft, cryptopunk, crypto |
| art | nft, nftcollector, cryptopunk |

specific groups and streams of communication. We identified 60,060 different hashtags and thus constructed a weighted undirected graph where: each node represents a hashtag and each connection between two nodes represents the presence of a tweet in which both corresponding hashtags were used. As the weight of each node was assigned the number of tweets in which the corresponding hashtag was present, and as the weights of each connection the number of tweets having the two corresponding hashtags in common. Figure 6a illustrates in descending order which are the 20 most used hashtags: `NFT`, `cryptopunks`, and `BAYC` are the most significant ones. Beyond the 20th, namely `art`, the weights are much lower. Overall, the properties of the graph thus formed are shown in Table 3a: the graph consists of 762,950 connections, with an average weight of about 29,634 common tweets. The average clustering coefficient and the relatively significant number of connected components (that is 439) suggest the presence of highly specialized communities within this content graph. However, edge density is extremely low in this case, being hashtags mainly connected with some central hashtags (see below). Most likely for this reason, modularity has a low value. While

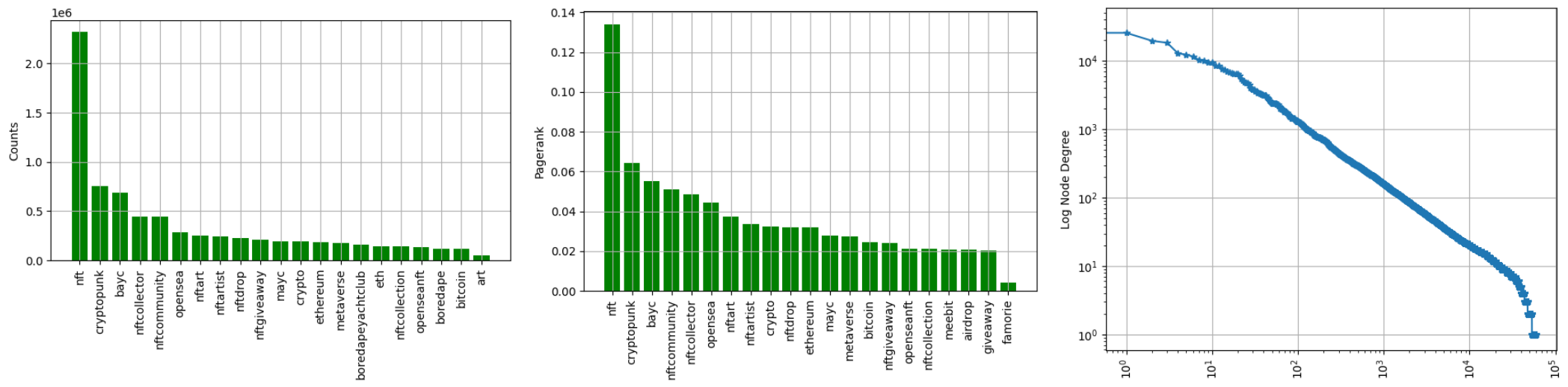

(a) Size (top 20 nodes) (b) Pagerank (top 20 nodes) (c) Connection degree distribution

Fig. 6. Size, pagerank and connection degree distributions of the hashtags graph illustrated in Table 3.

Figure 6b illustrates in descending order which are the 20 hashtags with highest pagerank. Again, the values beyond the 20-th (`famorie`, related to the CryptoMories) the weight is much lower. Figure 4b illustrates the graph taking into account only these 20 hashtags, where the size of the nodes is proportional to the pagerank: again, as can be observed, the hashtags `NFT`, `cryptopunks` and `BAYC` are the most influential. We also performed a commonality analysis for the hashtags: this was summarized in Table 3c. As can be observed, the NFT hashtag is the one that is most common to all.

The hashtags nftart, nftartist, and nftcollection are neither close to cryptopunks nor BAYC: this indicates how for more generic topics, and not only related to PFP, the discourses are less polarized on cryptopunks and BAYC.

### 4.4 The role of the social network community

Figure 7 shows trends in volume, average price, tweets, and users (wallets and Twitter accounts) for a number of very different projects from the date of creation until April 14, 2022. BAYC, CryptoHodlers, Mekaverse, and mfers were considered. As can be seen, the BAYC built up a continuous engagement on Twitter where within a year the number of unique users talking about them grew exponentially, as did the average price, which went from 0.08ETH to more than 100ETH in less than a year. There are projects, such as CryptoHodlers, on the other hand, where although the creators try to build an ecosystem around their project by distributing comics or developing a new reserved NFT, the community has not grown, impacting the average price and liquidity (in terms of daily trades) of this project. Mekaverse and mfers, summarize these concepts: for the former we see how a high number of tweets in the early stages of the project resulted in an initial high exchange value where the figurines traded for a minimum price of 8ETH, then went to 10 times less as the number of tweets dropped. For the latter, on the other hand, we see that engagement in terms of daily tweets coincides with an increase in average price, and this occurs gradually and constantly after their release.

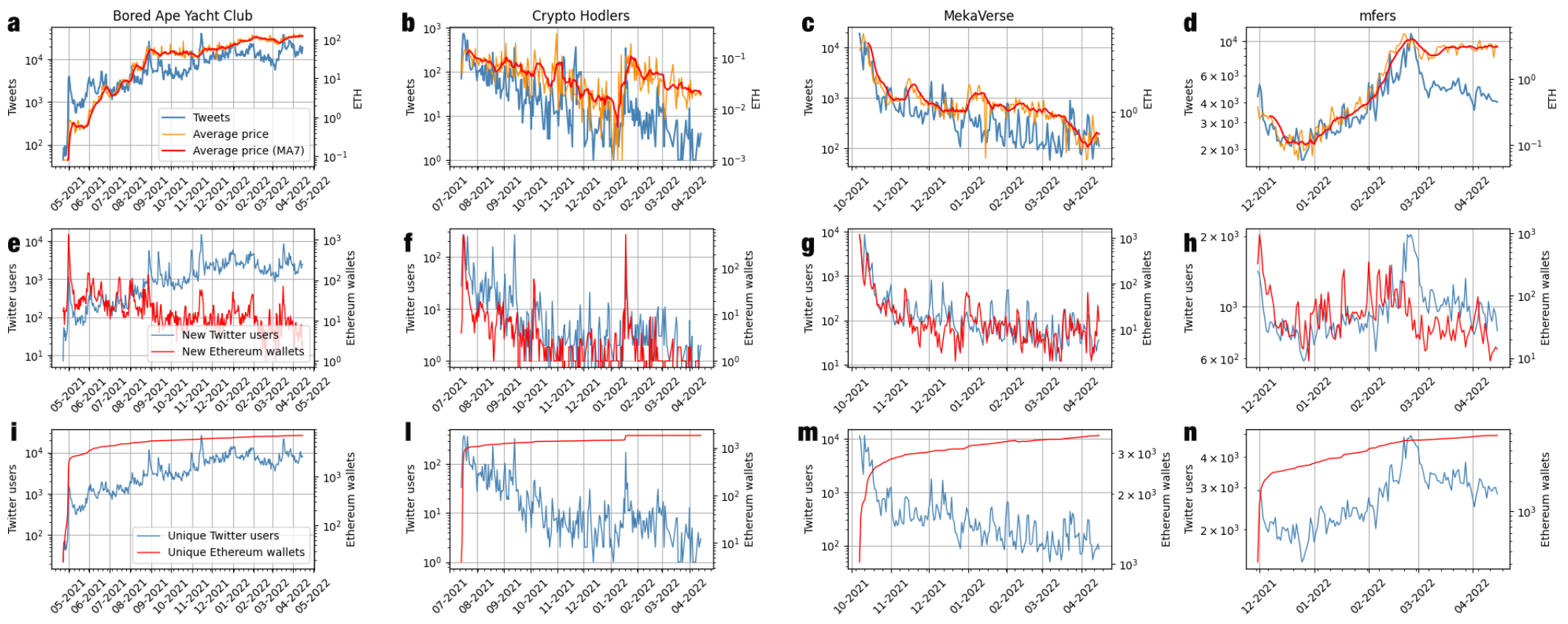

Fig. 7. (a-d) Daily number of tweets and average collection price in ETH, (e-h) daily new Twitter users and Ethereum wallets, (i-m) daily unique Twitter users and Ethereum wallets for the selected projects.

## 5 CONCLUSIONS

The NFT phenomenon is a topic that became mainstream during 2021 and has continued to receive increasing attention from early 2022 to the present. This blockchain-based technology has also come to the forefront thanks to projects such as BAYCs whose images have been purchased by celebrities and traded at very high prices as reported by several media. In this paper, we analyzed the phenomenon of NFTs, focusing on the class related to PFPs. We chose 18 projects based on our knowledge in the field and based on their importance in terms of volumes traded, users, and social activity. It was illustrated through analysis of social activity how the concept of social community is essential to ensure liquidity and value of PFP projects, i.e., used as social icons. The centrality and influence of the community related to BAYCs was demonstrated, that despite being second to CryptoPunks in terms of volumes traded, social activity is much higher and central to every other project in the industry. This work aimed to lay the groundwork for more in-depth future studies on understanding whether patterns and causality conditions may exist between elements such as the number of tweets, average price of a collection, Twitter users, and Ethereum wallet.